\documentclass[12pt]{article}

\usepackage{amsmath,amsfonts,amssymb,color,graphicx,overpic,}
\usepackage{slashed}
\usepackage{epsfig}
\usepackage[utf8]{inputenc}
\usepackage{amsmath}
\usepackage{ wasysym }
\usepackage{graphicx}
\usepackage[english]{babel}
\usepackage{graphic x}
\usepackage{relsize}
\usepackage{cite}

\begin{document}

\begin{titlepage}
\rightline{June 2018}
\vskip 1.9cm
\centerline{\large \bf
Shielding of a direct detection
experiment }
\vskip 0.3cm
\centerline{\large \bf
and implications for the DAMA annual modulation signal}

\vskip 2.4cm
\centerline{R. Foot\footnote{
E-mail address: rfoot@unimelb.edu.au}}

\vskip 0.4cm
\centerline{\it ARC Centre of Excellence for Particle Physics at the Terascale,}
\centerline{\it School of Physics, The University of Sydney, NSW 2006, Australia}
\vskip 0.1cm
\centerline{and}
\vskip 0.1cm
\centerline{\it ARC Centre of Excellence for Particle Physics at the Terascale,}
\centerline{\it School of Physics, The University of Melbourne, VIC 3010, Australia
}

\vskip 2.5cm
\noindent
Previous work has argued that, in the framework of plasma dark matter models, 
the DAMA annual modulation signal can be consistently explained with electron recoils.
In the specific case of mirror dark matter, that explanation requires an effective low velocity 
cutoff, $v_c \gtrsim 30,000$ km/s, for the halo mirror electron distribution at the detector.  We show here that  
this cutoff can result from collisional shielding of the detector from the halo wind
due to Earth-bound dark matter.
We also show that shielding effects can reconcile the
kinetic mixing parameter value inferred 
from direct detection experiments 
with the value favoured from small scale structure considerations, $\epsilon \approx 2 \times 10^{-10}$.


\end{titlepage}


For over a decade, the DAMA collaboration have observed an annually modulating rate
of scintillations in their NaI detector at Gran Sasso  \cite{dama1,dama2a,dama2b,dama3,dama4}.
This annual modulation signal has properties consistent with dark matter interactions, but since
there is no discrimination against electron recoils,
both nuclear and electron scattering
interpretations could be considered.
However, the nuclear option doesn't appear
to be possible in light of the rather stringent constraints from other experiments, e.g. \cite{xenon1T,panda,lux,cresst}.
In contrast, dark matter scattering off electrons is much less strongly constrained, and in  
fact, a consistent electron scattering interpretation of the DAMA annual modulation signal seems to be possible in 
the framework of plasma dark matter \cite{ef1,ef2,ef3}.
In that case
the DAMA signal results from `dark electron' scattering off atomic electrons 
in the NaI crystal facilitated by the kinetic mixing interaction. 
The average interaction rate is consistent with the results from other experiments, with the DarkSide-50 experiment \cite{darkside} providing
the most useful information.
That experiment sees a small excess at low energies which can be interpreted as dark matter induced electron
recoils, and
in combination with the results from DAMA, indicates that the annual modulation amplitude is near maximal.
This conclusion is supported by XENON100 results \cite{xene1,xene2} as analysed in \cite{ef3}.

The prototype plasma dark matter model posits the existence of a hidden sector containing
a dark electron, dark proton interacting with a massless dark photon, e.g. \cite{sunny}. This dark QED model additionally
features kinetic mixing with the ordinary photon \cite{fhe}, an interaction that imbues the dark proton and dark electron with 
tiny electric charges, $\pm \epsilon$ \cite{holdom}.
Since the dark electrons are electrically charged,
they can  scatter off atomic electrons via Coulomb scattering.

A simple analytic form for the cross section arises if
we approximate the target electron as free and at rest relative to the incoming dark
electron of speed $v$,
\begin{eqnarray}
\frac{d\sigma}{dE_R} = \frac{\lambda}{E_R^2 v^2}
\label{cs}
\end{eqnarray}
where $\lambda = 2\pi \epsilon^2 \alpha^2/m_e$,
and $E_R$ is the recoil energy of the scattered electron.
Treating the target electrons as free can only be approximately valid for the loosely bound atomic electrons,
i.e. those with binding energy much less than $E_R$.
We define $g_T(E_R)$ as the number of electrons per target atom with atomic binding energy ($E_b$) less than $E_R$, and we approximate
the electron scattering rate per target atom by replacing $\lambda \to g_T \lambda$ in Eq.(\ref{cs}).
[For the DAMA experiment, the `atom' is a NaI pair.]
Typically, the proportion of loosely bound electrons, i.e. with $E_b \ll E_R$, greatly outnumbers those with $E_b \sim E_R$, so
this approximation is expected to be reasonable.

The scattering rate of a dark electron off an electron is then:
\begin{eqnarray}
\frac{dR_e}{dE_R} &=&
N_T  n_{e'}
\int \frac{d\sigma}{dE_R}
\ f({\textbf{v}}; {\textbf{v}}_E; \theta)\ |{\textbf{v}}| \
d^3v \nonumber \\
&=& g_T N_T  n_{e'}
\frac{\lambda}{E_R^2}
\ I({\textbf{v}}_E,\theta)
\label{55}
\end{eqnarray}
where
\begin{eqnarray}
I({\textbf{v}}_E,\theta)
\equiv \int^{\infty}_{|{\bf{v}}| > v_{min} (E_R)}
\ \frac{f({\textbf{v}}; {\textbf{v}}_E; \theta)}{|{\textbf{v}}|} \ d^3 v
\ .
\label{III}
\end{eqnarray}
Here, $N_T$ is the number of target atoms per kg of detector,
$v_{\rm min} \ = \ \sqrt{E_R m_e/2\mu^2}$, where $\mu$ is the electron - dark electron reduced mass ($\mu = m_e/2$ for the mirror
dark matter case), and
$n_{e'}$ is the dark electron number density.
Also, $f({\textbf{v}}; {\textbf{v}}_E; \theta)$
is the velocity distribution of dark electrons which arrive at the detector.
As indicated, this distribution will depend on the velocity of the halo wind as measured from Earth
[${\textbf{v}}_E (t) $],  and 
the angle between the direction of the halo wind
and the zenith at the detector's location
[$\theta (t)$].
It might also depend on the detector's geographical location.

Expanding $n_{e'} I$ in a Taylor series around the (yearly) average, $\langle n_{e'} I \rangle \equiv n^0_{e'}/v_c^0$,
leads to a simple phenomenological model for the rate: 
\begin{eqnarray}
\frac{dR_e}{dE_R} =
g_T N_T  n^0_{e'}  \frac{\lambda}{v_c^0 E_R^2}\left[ 1 + A_v\cos\omega (t-t_0) + A_\theta (\theta - \bar \theta)\right]
\ .
\label{r68}
\end{eqnarray}
If $\langle |\textbf{v}|\rangle \gg v_{\rm min}$,
then the velocity integral, $I({\textbf{v}}_E,\theta)$, and hence also $v_c^0$,  becomes approximately independent of $E_R$.
Since $v_{\rm min} \propto \sqrt{E_R}$, this regime could only occur for sufficiently small $E_R$, below some threshold, $E_R^T$.
For $E_R > E_R^T$ the rate becomes strongly suppressed.

The above rate [Eq.(\ref{r68})] was found in \cite{ef3} to be compatible with both the DAMA annual modulation signal and the 
low average rate observed by the DarkSide-50 and XENON100 collaborations,
provided that $E_R^T \gtrsim 2$ keV and
the annual modulation is near maximal, $A_v \approx 1$.
In addition,
the normalization is approximately fixed by the data, leading to an estimate of $n_{e'}^0 \epsilon^2/v_c^0$, equivalent to:
\begin{eqnarray}
\epsilon 
\approx 1.5 \times 10^{-11} {\cal F}
\label{fix}
\end{eqnarray}
where
\begin{eqnarray}
{\cal F} = \sqrt{\frac{v_c^0}{50,000 \ {\rm km/s}}}
\sqrt{\frac{0.2 \ {\rm cm^{-3}}}
{n^0_{e'}}}
\ .
\label{f1}
\end{eqnarray}
Although not currently required to explain the experiments, $A_\theta$ can be nonzero, and probed via diurnal
variation.

In this paper we focus on the theoretically constrained mirror dark matter model 
(for a review and detailed bibliography, see \cite{footrev}),
where dark matter results from a hidden sector exactly isomorphic to the Standard Model \cite{flv}. 
In that case, the dark electron is the `mirror electron', the mass-degenerate partner of the electron,
and there is a spectrum of mirror nuclei, H$'$, He$'$, O$'$, Fe$'$, etc.
For mirror dark matter, 
$\epsilon \gtrsim 10^{-10}$ is required from halo dynamics; 
this limit assumes that
type II supernovae are the dark sector heat sources, required to dynamically balance radiative cooling losses within galaxies \cite{sph,footfeb,footapril}. 
Furthermore, $\epsilon \approx 2 \times 10^{-10}$
follows from the observed flat galaxy velocity function \cite{sunny2,footapril}.
With $\epsilon \gtrsim 10^{-10}$,
Eq.(\ref{fix}) suggests that either $v_c^0 \gg 50,000$ km/s and/or $n_{e'} \ll 0.2\ {\rm cm^{-3}}$. 

In this kind of dark matter model, dark matter is captured and accumulates within the Earth, where it thermalizes with the ordinary
matter and forms an extended distribution.
The upper layers of the Earth-bound dark matter can be partially ionized leading to the formation of a `dark ionosphere'.
The ionization of this layer would be due to the interactions of the halo mirror electrons with the Earth-bound dark matter.
A consequence of such a conducting layer is that induced dark electromagnetic fields can be generated which
deflect the halo wind. This situation is analogous to the way Venus and Mars deflect the solar wind (e.g. \cite{venus,venus2}), and has
been studied for the dark matter case in \cite{ef2} by numerically solving the fluid equations.
In addition, there can be collisional shielding of the detector if the Earth-bound dark matter extends to the Earth's
surface.  
These effects can 
not only  provide a mechanism for reducing $n_{e'}$, but 
can also suppress the flux of halo dark matter matter particles below some cutoff, $v_{\rm cut}$.
Although it is unclear which effect dominates, for the purposes of this paper we shall
focus on the collisional shielding, and provide only a crude estimate of this effect.

Dark matter that is bound within the Earth has been studied 
in the mirror matter case \cite{footdi}, more generally 
\cite{sunny3}, and also
recently in the context of a somewhat different kind of model with strong dark matter-baryon interactions \cite{grecent}.
We shall assume a spherically symmetric distribution, governed by the hydrostatic equilibrium condition:
\begin{eqnarray}
\frac{dP}{dr} = -\rho g
\end{eqnarray}
where $\rho$ is the dark matter density, $P$ its pressure, and $g$ the local acceleration due to the Earth's gravity.
Using $P = n_{A'}T$, $\rho = n_{A'} m_{A'}$, we have
\begin{eqnarray}
\frac{dn_{A'}}{dr} = -\frac{n_{A'}}{T}\left[m_{A'} g + \frac{dT}{dr}\right]
\ .
\label{na}
\end{eqnarray}
Within the Earth, i.e. $r \le R_E$, where 
$R_E \simeq 6371.0$ km is the radius of the Earth,
the mirror particles rapidly thermalize with the ordinary matter. The relevant distance scale can easily be estimated.
For $\epsilon \approx 2\times 10^{-10}$, a mirror atom with speed less than a few kilometers per second 
is thermalized with the ordinary environment after travelling less than a kilometer cf.\cite{foot2004}.
This means that the temperature distribution of the mirror particles within the Earth, approximately matches that of the ordinary matter.

The earlier study \cite{footdi} used Eq.(\ref{na}), with the mirror particle temperature distribution thermalized with
that of the ordinary matter,  to estimate the mirror particle distributions within the Earth.
That work suggests that this distribution can only extend to the surface of the Earth if $A'$ is sufficiently
light ($\lesssim 5$ GeV). Here we shall focus on the case where $A' = {\rm He'}$, so that $m_{A'} = 3.73$ GeV.
The mirror helium is expected to be bound into neutral atoms, except near the surface where interactions with the halo dark matter can 
lead to some level of ionization. The self interaction cross section of helium atoms is around $10^{-16}\ {\rm cm^2}$,
and the scattering length is around $[n_{\rm He'} \sigma]^{-1} \approx 0.1 [10^{12}\ {\rm cm^{-3}}/n_{\rm He'}] \ {\rm km}$.

The temperature profile of the He$'$ in the
Earth's atmosphere ($R_E \lesssim r  \lesssim R_E + 100 \ {\rm km}$) is an important quantity of interest.
The ordinary matter density varies by around six orders of magnitude over this 100 km
height range. In the upper atmosphere, the interaction rate of the captured He$'$ with ordinary matter becomes
too low to thermalize
the He$'$ distribution with that of the ordinary matter. The expected result is a temperature profile which initially follows that of the ordinary
matter at lower altitudes, but at higher altitudes sharply rises; the exact transition point could be determined by careful modelling of the heating and
cooling processes.
For the purposes of a rough estimation, we ignore this complexity and
adopt a simple isothermal model for the mirror helium temperature in the Earth's atmosphere, taking $T=240$ K.
Considering the radial region extending up to a  hundred kilometers or so above the Earth's surface,
$g$ is approximately constant,  $g\simeq 9.8 \ {\rm m/s}^2$, and
Eq.(\ref{na}) has the analytic solution:
\begin{eqnarray}
n_{\rm He'}(r) = n_{\rm He'}(R_E) e^{-\lambda_s (r-R_E)} \ \ \ {\rm for \ r > R_E}
\ .
\label{x8}
\end{eqnarray}
Here $\lambda_s = m_{\rm He'} g/T \approx (50 {\rm km})^{-1}$ for $T=240$ K.
In the case where $r < R_E$, the density can be obtained by numerically solving Eq.(\ref{na}).


A halo mirror electron loses energy due to collisions with the Earth-bound mirror helium atoms.
The energy loss per distance $d\ell$ travelled is:
\begin{eqnarray}
\frac{dE'}{d\ell} = -n_{He'} \int_{E_{\rm min}}^{E'} \frac{d\sigma}{dE_R} \ E_R dE_R
\label{x9}
\end{eqnarray}
where the cross section has the form Eq.(\ref{cs}), with $\lambda = g_T 2\pi \alpha^2/m_e$.
For mirror electron interactions with neutral He$'$ atoms, $g_T = 2$ for $E_R >  E_b$,
where $E_b \simeq 25$ eV is the atomic binding energy for helium.
For $E_R <  E_b$, the energy is insufficient to ionization helium, and we can take $E_{\rm min} = E_b$.
Since $d\sigma/dE_R \propto 1/v^2$, the distance between collisions goes like $v^2$, and so for $v$ sufficiently large,
the effects of collisions become negligible.
For the purposes of this rough estimation, we shall approximate this behaviour as a sharp transition.
That is, for $v \gtrsim v_{\rm cut}$, collisions can be neglected, while for $v \lesssim v_{\rm cut}$ collisions are
important, and effectively slow the halo electrons until they have energy $\sim E_b \approx 25$ eV. 
The transition velocity can be easily estimated from Eq.(\ref{x9}), and is given by:
\begin{eqnarray}
v_{\rm cut}^4 \approx \frac{16\pi}{m_e^2} \alpha^2 \Sigma \log \Lambda 
\label{cut}
\end{eqnarray}
where $\Lambda \sim T/E_{\rm min}\approx 20$ 
and $\Sigma = \int n_{\rm He'} d\ell$.

The transition velocity, $v_{\rm cut}$,
depends on the direction of the incoming mirror electron.
The optical depth for mirror electrons 
coming vertically down will be lower, and a lower $v_{\rm cut}$
will result compared with mirror electrons coming from the horizon.
Approximating the path of the energetic halo mirror electrons as a straight line, and using Eq.(\ref{x8}) for the density, we find
the column density:
\begin{eqnarray}
\Sigma (\psi )  = n_{\rm He'}(R_E) \int e^{-\lambda_s d(\psi)} d\ell
\ ,  \ \ {\rm for} \ \frac{\pi}{2} \le \psi \le  \pi
\label{12b}
\end{eqnarray}
where $d(\psi) \equiv \sqrt{\ell^2 + R_E^2 - 2\ell R_E \cos\psi}$.
Here, $\psi$ is the angle between the direction of the incoming mirror electron and the zenith at the detector's location.
In this notation, $\psi = \pi$ corresponds to a particle travelling vertically down; the direction for which the column density will be a minimum.

To model the effect of this cutoff, we shall assume
that the mirror electron distribution is Maxwellian far from the Earth. That is, $f(v) = exp(-E/T)/k = exp(-v^2/v_0^2)/k$, 
where $v_0 = \sqrt{2T/m_e} \approx 11,100 \sqrt{T/0.35 \ {\rm keV}}$ km/s and $k \equiv v_0^3\pi^{3/2}$.
We shall first consider (yearly) averaged quantities, which we model for now with a time-independent $v_0$.
Under these assumptions, 
\begin{eqnarray}
\frac{1}{v_c^0} \equiv \langle I \rangle 
&\equiv & \frac{1}{{\cal N}} \int^{\infty}_{|{\bf{v}}| > y}
\ 
\frac{e^{-v^2/v_0^2}}{v_0^3 \pi^{3/2} |\bf{v}|} 
 \ d^3 v
\nonumber \\
& = & \frac{1}{{\cal N} v_0\sqrt{\pi}} \int e^{-y^2} \ d\cos\psi
\label{IIIb}
\end{eqnarray}
where 
$y \equiv MAX [v_{\rm cut}(\psi), v_{\rm min}(E_R)]$ and ${\cal N}$ is the normalization factor 
\begin{eqnarray}
{\cal N}  = \int^{\infty}_{|{\bf{v}}| > v_{\rm cut}(\psi)}
\frac{e^{-v^2/v_0^2}}{v_0^3 \pi^{3/2}} 
 \ d^3 v
\ .
\end{eqnarray}
The number density of the surviving high velocity component arriving 
at the detector is related to the density far from the Earth via:
\begin{eqnarray}
n_{e'}^0 = {\cal N} n_{e'}^{\rm far}
\ .
\label{15}
\end{eqnarray}
In the numerical work we take $n_{e'}^{\rm far} = 0.2\ {\rm cm^{-3}}$.

\begin{figure}[t]
  \begin{minipage}[b]{0.5\linewidth}
    \centering
    \includegraphics[width=0.7\linewidth,angle=270]{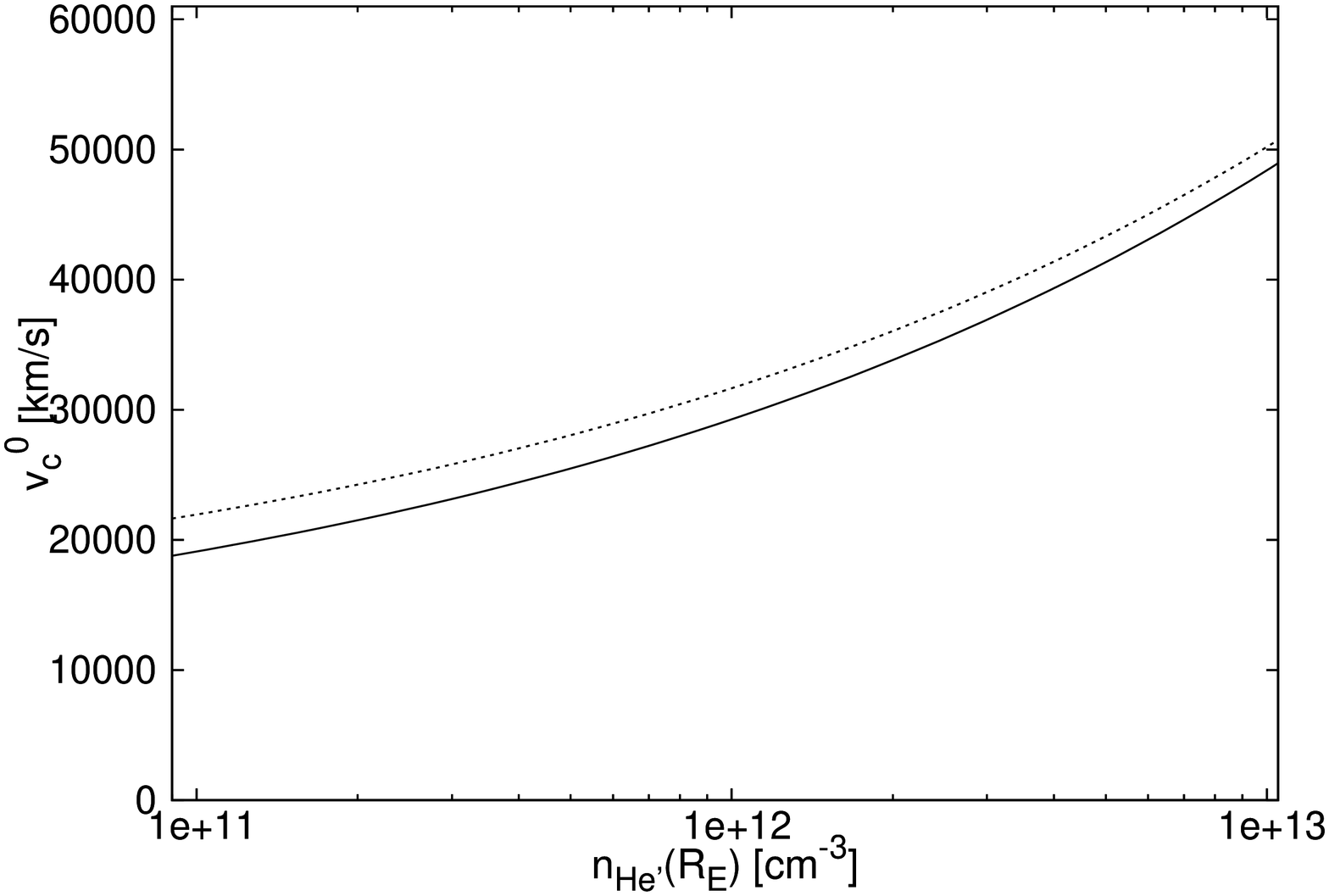}
     (a)
    \vspace{4ex}
  \end{minipage}
  \begin{minipage}[b]{0.5\linewidth}
    \centering
    \includegraphics[width=0.7\linewidth,angle=270]{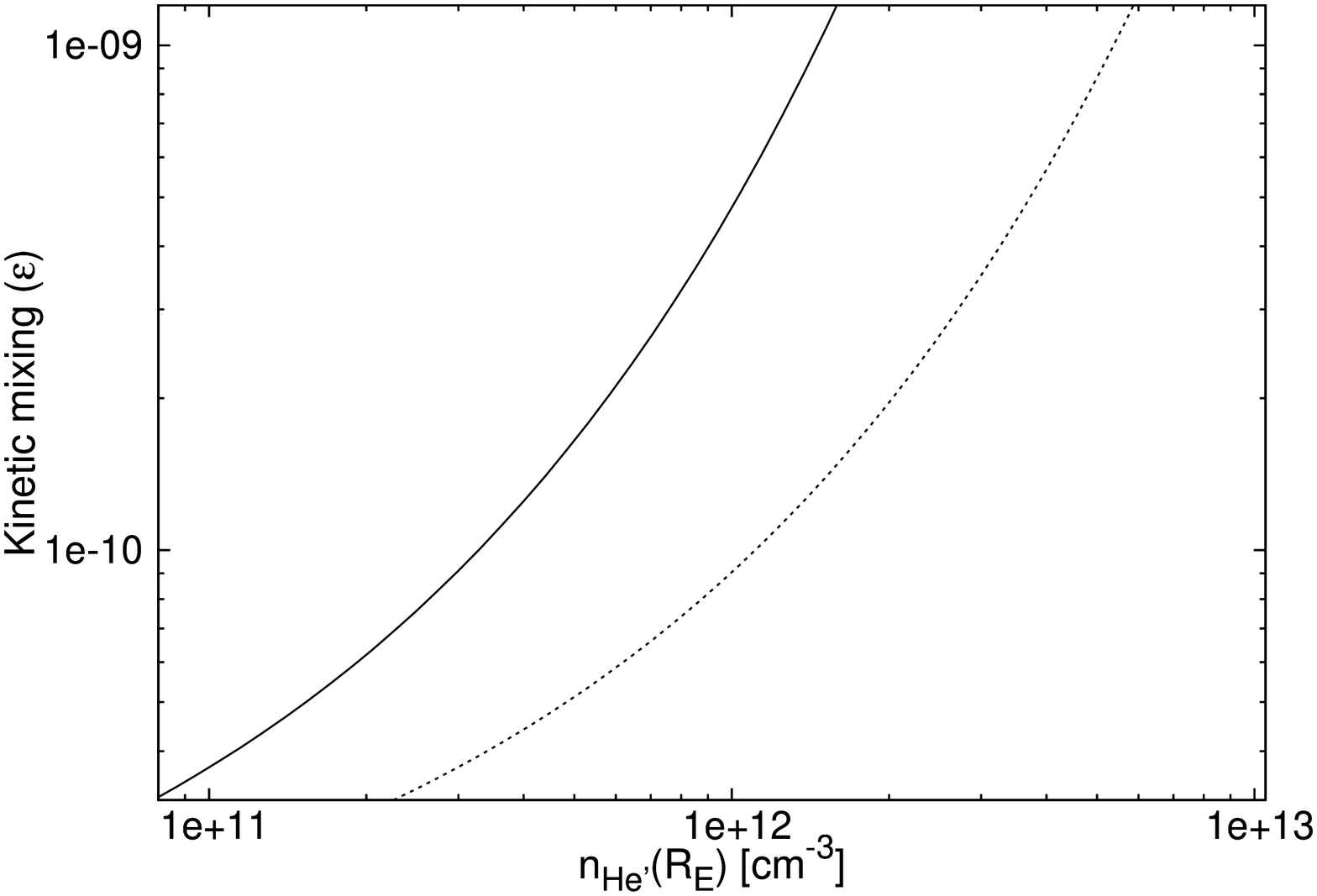}
    (b)
    \vspace{4ex}
  \end{minipage}
\vskip -1.0cm
\caption{
\small
(a) The quantity $v_c^0 \equiv 1/\langle I \rangle$ calculated from Eq.(\ref{IIIb}), and (b) the kinetic mixing parameter,
calculated from Eqs.(\ref{IIIb},\ref{15},\ref{fix}),
each plotted in terms of the 
number density of mirror helium atoms at the Earth's surface, $n_{\rm He'}(R_E)$.
The solid line (dashed line) shows the results for $T=0.3$ keV ($T=0.6$ keV).
}
\end{figure}

By numerically integrating Eq.(\ref{IIIb},\ref{12b},\ref{cut}), we can estimate $n_{e'}^0$, and $v_c^0$ in terms of the number density of mirror helium
at the Earth's surface, $n_{\rm He'}(R_E)$. 
The results for $v_c^0$ are shown in Figure 1a for two representative values for the halo temperature far from the Earth, $T=0.30$ keV, $T=0.60$ keV.
We can also estimate the quantity, ${\cal F}$, [Eq.(\ref{f1})],  and given Eq.(\ref{fix}), also the kinetic mixing strength, $\epsilon$.
The results for the kinetic mixing strength is shown in Figure 1b for the same two representative $T$ values.
In each case we have assumed that the recoil energies are low enough so that $v_{\rm min}(E_R) < v_{\rm cut}(\theta)$. 

From Fig.1b we see that for $\epsilon = 2\times 10^{-10}$, $n_{\rm He'}(R_E) \sim 10^{12} \ {\rm cm^{-3}}$.
This density is consistent with the rough estimations given in Ref.\cite{footdi}.
Indeed, He$'$ is expected to be captured at a rate:
\begin{eqnarray}
R_{\rm cap} &=& f_{\rm abs} \pi R_\oplus^2 v_{\rm rot} n_{\rm He'} \nonumber \\
            &\approx & f_{\rm abs} 10^{32} \left( \frac{n_{\rm He'}}{0.1 \ {\rm cm^{-3}}}\right) \ {\rm yr^{-1}} \ 
\label{cap}
\end{eqnarray}
where $v_{\rm rot} \approx 230$ km/s is taken as the average He$'$ speed
in the Earth frame.
The above estimation includes a He$'$ absorption fraction,
$f_{\rm abs}$, as
a significant fraction of He$'$ can be deflected due to induced dark electromagnetic fields in the dark ionosphere.
During $5$ Gyr, Eq.(\ref{cap}) indicates that the total number of captured He$'$ is around $5 \times 10^{41}\ f_{\rm abs}$. Using the
Earth baryonic density
and temperature profiles (Fig.2) \cite{Earthmodel} 
numerical solution of Eq.(\ref{na}) yields a number density 
of around $n_{\rm He'}(R_E) \sim f_{\rm abs} \ 10^{13}\ {\rm cm^{-3}}$,
shown in Figure 3.

\begin{figure}[t]
  \begin{minipage}[b]{0.5\linewidth}
    \centering
    \includegraphics[width=0.7\linewidth,angle=270]{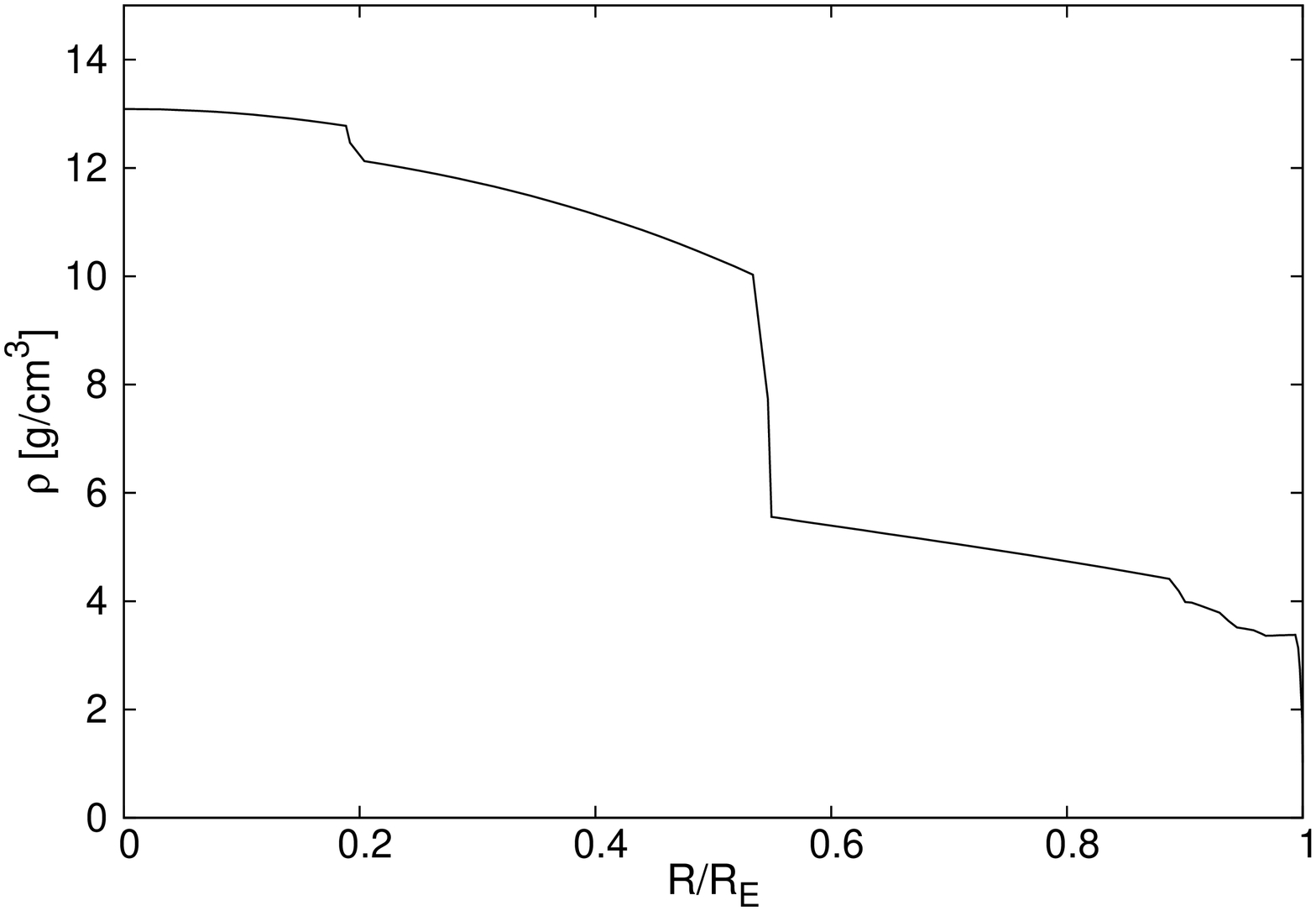}
     (a)
    \vspace{4ex}
  \end{minipage}
  \begin{minipage}[b]{0.5\linewidth}
    \centering
    \includegraphics[width=0.7\linewidth,angle=270]{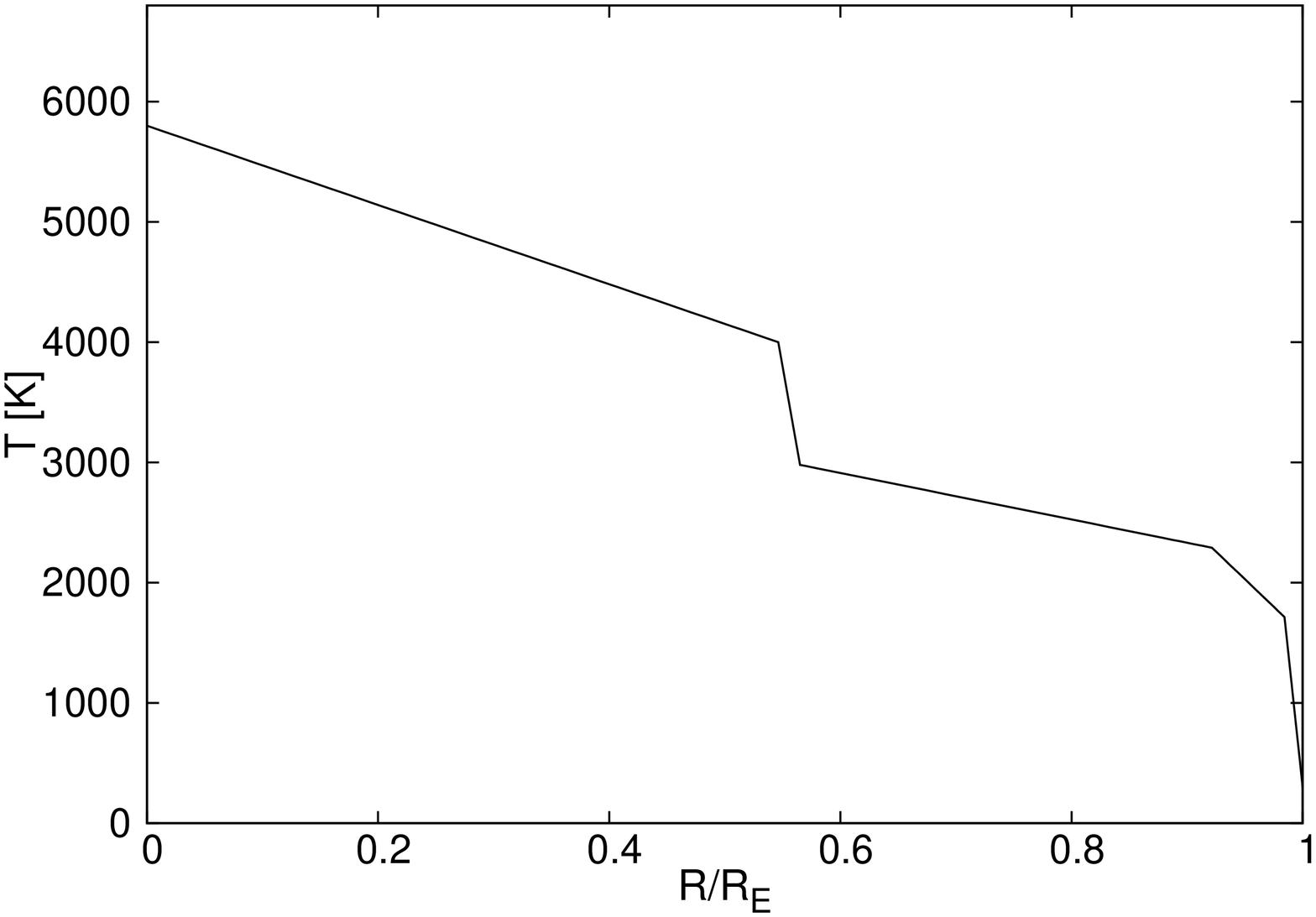}
    (b)
    \vspace{4ex}
  \end{minipage}
\vskip -1.0cm
\caption{
\small
The baryonic density (a) and temperature (b) profiles within the Earth.
}
\end{figure}

In Ref.\cite{footdi} it was suggested that the captured He$'$ might ultimately escape due to evaporation, as a significant fraction
can have velocity larger than the Earth's escape velocity, $v_{\rm esc} \approx 11.2$ km/s.
This, of course, depends on the temperature of the distribution near the last scattering surface (LSS).
If we assume an isothermal distribution at temperature $T_{\rm LSS}$, with number density $n_{\rm LSS}$, then
the flux of escaping He$'$ is given by the Jean's formula
\cite{Jean}
\begin{eqnarray}
F_{\rm esc} = \frac{n_{\rm LSS} v_{\rm LSS}}{2\sqrt{\pi}}\left(1 + \frac{v_{\rm esc}^2}{v^2_{\rm LSS}}\right) \exp\left(-v_{\rm esc}^2/v^2_{\rm LSS}\right)
\end{eqnarray}
where $v_{\rm LSS} = \sqrt{2T_{\rm LSS}/m_{\rm He'}}$.
For neutral He$'$-He$'$ collisions, the elastic scattering cross section is $\sigma \approx 10^{-16} \ {\rm cm^2}$, and
$n_{\rm LSS} \sim \lambda_s/\sigma \sim 10^9\ {\rm cm^{-3}}$.
The thermal evaporation rate, $4\pi R_\oplus^2 F_{\rm esc}$, is less than the capture rate, Eq.(\ref{cap}),  provided $T_{\rm LSS} \lesssim 1400$ K.
It appears that an extended He$'$ distribution is possible.
Interestingly, evaporation effects can potentially provide a mechanism to limit the amount of He$'$ particles
gravitationally bound to the Earth.
In particular, if the He$'$ evaporation
rate happened to be greater than the capture rate, then the consequent reduction of He$'$ particles would result in the lowering of the altitude of the LSS.
In the realistic case, where the He$'$ temperature profile is a rising function of altitude, evaporation can occur until the altitude of the LSS
is low enough so that $T_{\rm LSS} \approx 1400$ K. That is, the system would dynamically adjust until the evaporation and capture rates balance.
A detailed investigation along these lines might yield an estimate for the density profile, fixing the value for the number density of He$'$ at $R_E$, 
$n_{\rm He'}(R_E)$,
although for now we leave this as an uncertain parameter.

\begin{figure}[t]
    \centering
    \includegraphics[width=0.48\linewidth,angle=270]{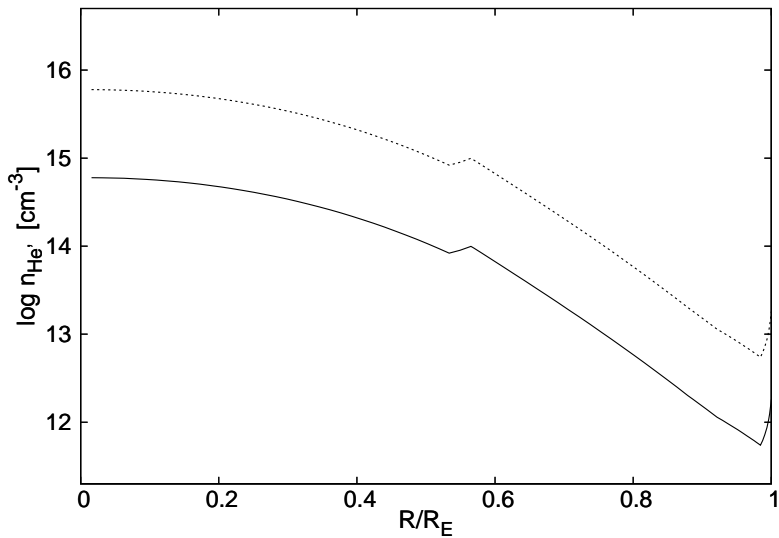}
\caption{
\small
Number density of He$'$ within the Earth.
The dashed line assumes that all He$'$ particles within the Earth's path get absorbed into the Earth,
while the solid line assumes only $10\%$ of these particles get absorbed.
}
\end{figure}

Consider now the time-dependence of the electron interaction rate.
A large time variation is expected if the Earth-bound dark matter has a partially ionized surface layer, a dark ionosphere.
Such a partially ionized layer arises due to the ionizing
interactions of halo dark matter at $r \approx R_E$.\footnote{The free $e'$ density can be estimated by balancing the He$'$ ionization
rate ($e' + {\rm HeI} \to {\rm HeII} + e' + e'$)  
with the capture rate of He$'$ ions ($e' + {\rm HeII} \to {\rm HeI} + \gamma'$). 
A rough calculation along these lines suggests a free $e'$ density of around $\sim 10^4 \ {\rm cm^{-3}}$ for
$n_{\rm He'} = 10^{12} \ {\rm cm^{-3}}$, which is typical of the electron density that arises in the ionosphere of the planets.}
Modelling the halo dark matter as a fluid, the earlier study \cite{ef2} numerically solved the fluid equations
and found significant
annual (and diurnal) temperature and density variations
of the modelled halo dark matter distribution near the dark ionosphere.
In fact, since
the distribution is cutoff, with only the high velocity tail possibly able to reach
the detector, the
interaction rate can be exponentially sensitive to variations in the temperature.

Of course, with the low velocity cutoff the distribution is not Maxwellian, but
modelled via Eqs.(\ref{IIIb},\ref{12b},\ref{cut}).
The variations of temperature as described by fluid equations will translate into variations of 
the velocity dispersion, $v_0 = \sqrt{2T/m_e}$.
We explore the time dependence by considering the effect on $v_0$ and hence 
$I({\textbf{v}}_E,\theta)$
from a temperature variation
\begin{eqnarray}
T = T_0 + T_v \cos\omega (t-t_0)
\end{eqnarray}
where $\omega = 2\pi/{\rm year}$, $t_0 \simeq 153$ days (June 2$^{nd}$).
In Figure 4, we show the yearly average, modulation amplitude, and time variation of the rate for the DAMA experiment (NaI 
detector with energy resolution as measured by DAMA \cite{res}).
Two representative temperature values are considered,
$T_0 = 0.3 \ {\rm keV}$, $T_v = 0.05$ keV, and 
$T_0 = 0.6 \ {\rm keV}$, $T_v = 0.1$ keV.
In each case the (assumed time-independent) $n_{\rm He'}(R_E)$ value adopted was obtained from 
Figure 1b assuming $\epsilon = 2\times 10^{-10}$, i.e. $n_{\rm He'}(R_E) = 5.8\times 10^{11}\ {\rm cm^{-3}}$ 
[$2.0\times 10^{12} \ {\rm cm^{-3}}$] for $T_0/keV = 0.3$ [$0.6$].
As indicated in Figure 4b,  the rate modulation is nearly maximal ($A_v \simeq 0.7$), and the modulation is also close
to sinusoidal.
The numerical study \cite{ef2} found even larger temperature variations near the dark ionosphere, so a near maximal
annual modulation amplitude appears to be plausible provided that the altitude of the dark ionosphere is below/near
that of the detector.

\begin{figure}[t]
  \begin{minipage}[b]{0.5\linewidth}
    \centering
    \includegraphics[width=0.7\linewidth,angle=270]{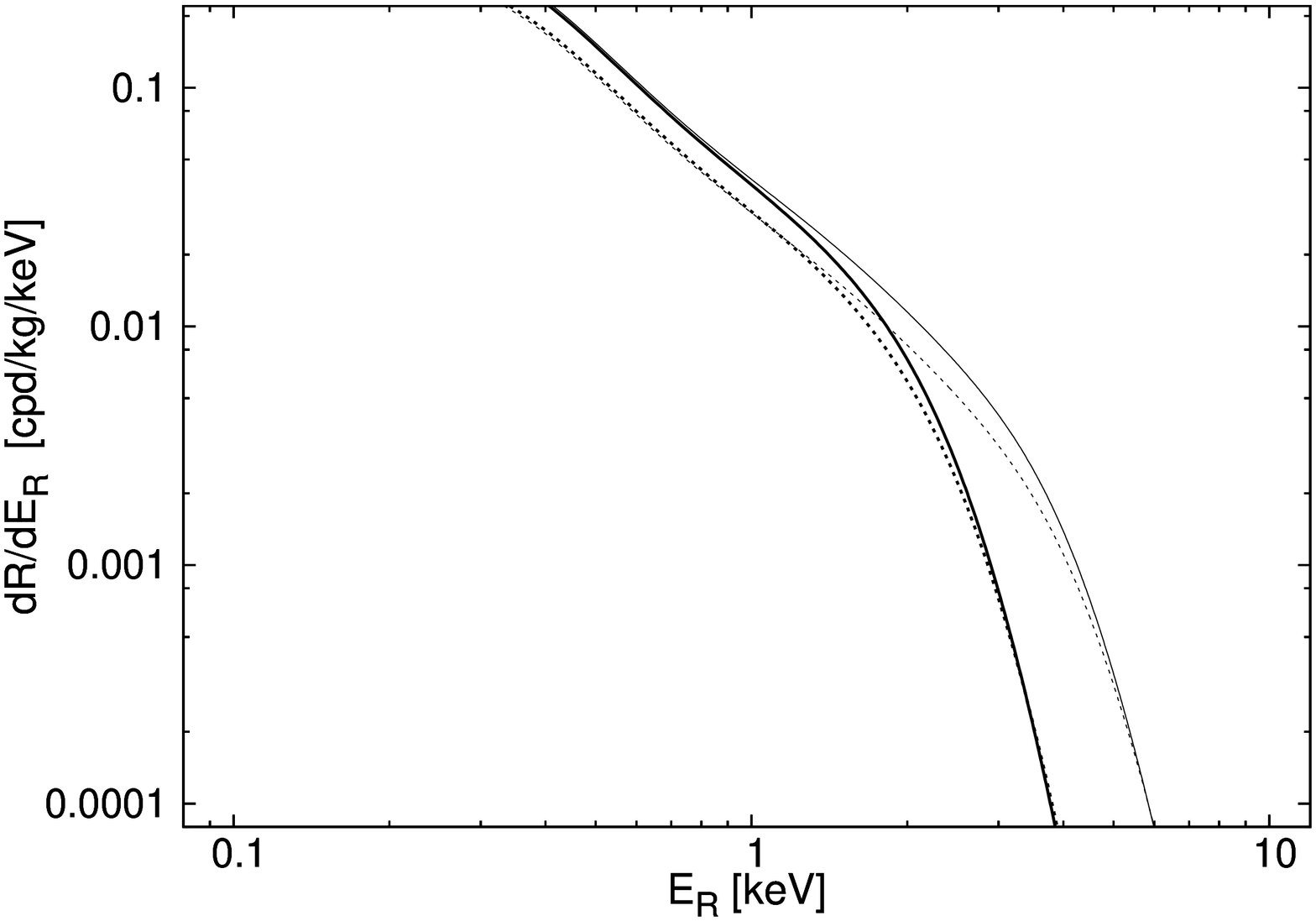}
     (a)
    \vspace{4ex}
  \end{minipage}
  \begin{minipage}[b]{0.5\linewidth}
    \centering
    \includegraphics[width=0.7\linewidth,angle=270]{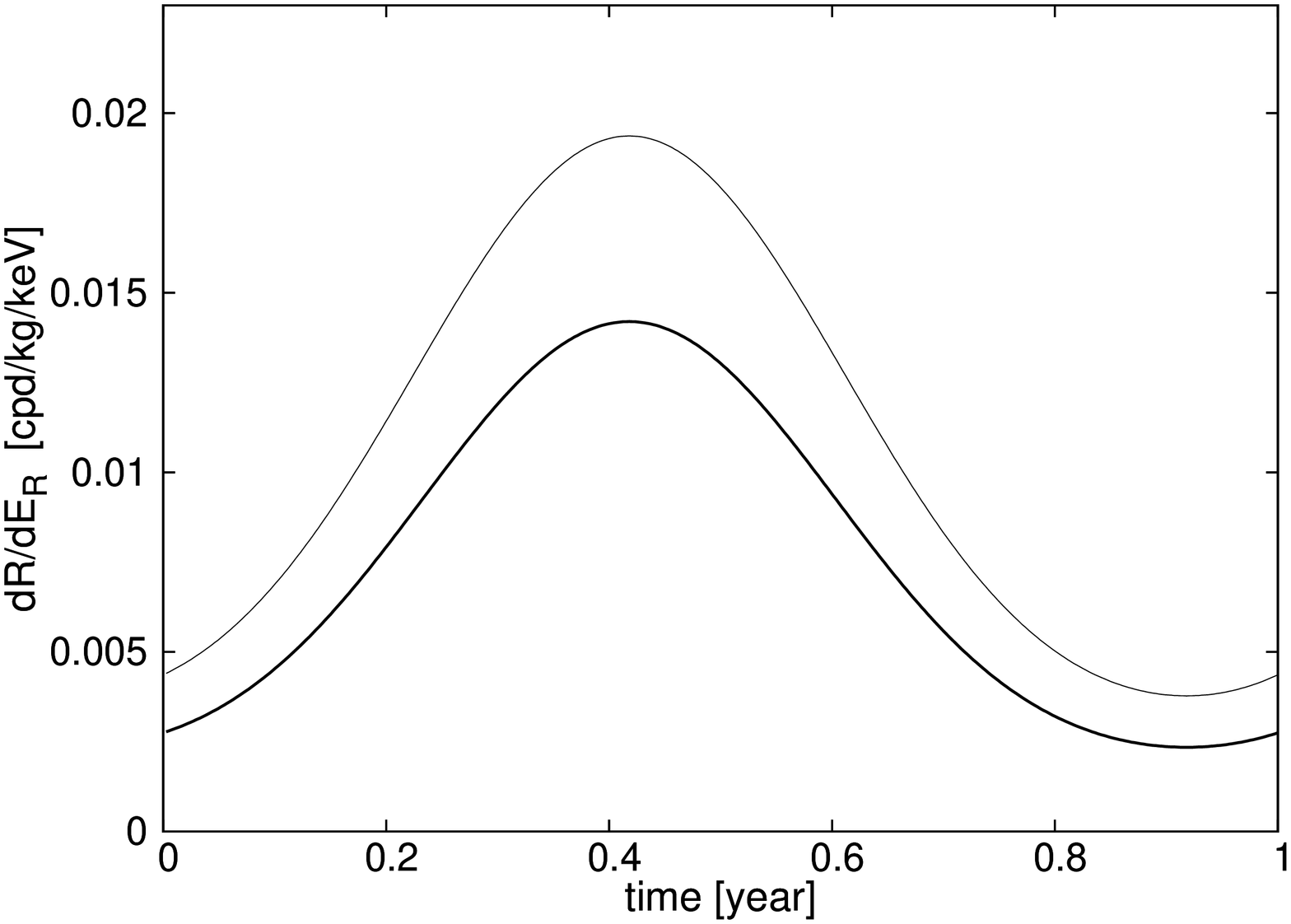}
    (b)
    \vspace{4ex}
  \end{minipage}
\vskip -1.0cm
\caption{
\small
(a) Predicted average rate (solid line) and annual modulation amplitude (dashed line)
for the DAMA set up.
The thick lines assume 
$T_0 = 0.3$ keV, $T_v = 0.05$ keV, while the thin lines are for
$T_0 = 0.6$ keV, $T_v = 0.1$ keV.
(b) The time variation of the rate averaged over the recoil energy range $1 \le E_R/{\rm keV} \le 4$, for the
same parameters as panel (a).
}
\end{figure}

Figure 4 demonstrates that the rate of $E_R \gtrsim 2$ keV electron recoils
can be strongly suppressed. This is due to the onset of the threshold at $E_T \sim m_e (v_c^0)^2/2$.
For experiments such as XENON100, XENON1T, LUX etc., there is a loss of detection efficiency in the S1 signal which happens
to coincide (roughly) with this energy scale,
and consequently the expected rates in those experiments can be very low.
Such experiments, unfortunately, may not be able to conclusively confirm or exclude the DAMA signal.
Indeed, for the same parameters as adopted in Fig.4,
we have estimated the rate for the XENON1T experiment, 
assuming a sharp cutoff in detection efficiency at 2.2 keV.
(For the purposes of this exercise, the resolution was assumed to be identical to that of the
DAMA experiment.)
The result is a yearly average rate in the 2-6 keV energy range of 
$1.0\times 10^{-4}$ cpd/kg/keV for  $T_0 = 0.3 \ {\rm keV}$ and $1.5\times 10^{-3}$ cpd/kg/keV for $T_0 = 0.6 \ {\rm keV}$.
These values can be compared with the rate of electron recoils reported by XENON1T of $2\times 10^{-4}$ cpd/kg/keV \cite{xenon6655} over the same energy range.
This measurement would appear to disfavour the $T_0 = 0.6$ keV benchmark point, although
the level of tension might not be so severe given that
the XENON1T events reported in \cite{xenon6655} were collected during 34.2
live days between Nov 22, 2016 - Jan 18, 2017. This period happens to
approximately coincide with Dec.1$^{st}$, the time at which
the rate is at the yearly minimum. Naturally, this minimum rate can be very low 
if the annual modulation is near maximal.

DarkSide-50, XENON1T, XENON100 and the DAMA experiments are all located in the Gran Sasso Laboratory, i.e. at the same geographical location. 
The halo $e'$ distribution should be virtually identical in each detector. It follows that these experiments can be compared with each other. 
However, the distribution of $e'$ at different geographical locations, need not be the same.
The distribution can potentially depend sensitively on modest variations of the Earth bound dark matter,
density and temperature.
The location of the detector relative to the 
dark ionosphere can vary at  different locations.
There may even be some dependence on the atmospheric temperature variations of the ordinary 
matter.
Of course, it is unclear if such   
effects can account for the intriguing results obtained at other locations, including the hint of a small annual
modulation, with amplitude of opposite sign to DAMA observed
in the Kamioka Observatory (XMASS)\cite{XMASS}  and in the Homestake mine (LUX) \cite{LUX}, but it might well be possible.


To conclude, we have focused on the theoretically constrained mirror dark matter model, and examined 
collisional shielding of direct detection experiments.
In particular, we have found that
Earth-bound mirror dark matter
can partially shield detectors from halo dark matter, as
the He$'$ distribution extends to the Earth's surface with density estimated to be around $\sim 10^{12}\ {\rm cm^{-3}}$.
This shielding can be effective for halo mirror electrons with sufficiently low velocity, while higher velocity
particles can reach the detector unimpeded. The transition velocity was estimated to
be around 30,000 km/s. 
In addition to collisional shielding, the surface layers of the Earth-bound dark matter are expected to be
partially ionized, leading to the formation of a dark ionosphere near the Earth's surface. Induced dark electromagnetic
fields in this layer can deflect the halo wind, which can also effectively shield detectors from the halo wind.

These shielding effects can be quite important. 
Previous work has found that the DAMA annual modulation signal can be consistently
explained with electron recoils. Within the mirror dark matter model, 
this interpretation requires an effective low velocity cutoff, $v_c \gtrsim 30,000$ km/s. The shielding effects considered here provide an explanation for
such a cutoff.
Furthermore, 
with shielding effects included,
the kinetic mixing strength inferred 
from direct detection experiments 
becomes compatible with the value
favoured from small scale structure considerations, $\epsilon \approx 2 \times 10^{-10}$.

\vskip 0.8cm
\noindent
{\large \bf Acknowledgments}

\vskip 0.3cm
\noindent
This work was supported by the Australian Research Council.

\end{document}